
\input phyzzx.tex

%
\pubtype{NCKU-HEP/93-06}
\titlepage

\title{SPONTANEOUS CP-VIOLATION IN THE EXTENDED STANDARD MODELS}

\author {Chilong Lin, Chien-er Lee, and Yeou-Wei Yang}
\address{Department of Physics, National Cheng Kung University, Tainan, Taiwan,
Republic of China}

\abstract
 The spontaneous CP-violation in three kinds of extension of the standard
model are investigated. We find that if no additional symmetry is introduced to
reduce the number of coefficients, then all three extended models possess
the spontaneous CP violation (SCPV) in the
Higgs potential. When symmetry under which the doublets transform as
$\Phi_1 \to \Phi_1$ and $\Phi_2 \to e^{i\gamma}\Phi_2$ is introduced,
the two-Higgs doublet model will lose its SCPV terms.
In the other two extended models, the SCPV terms will be preserved since both
models contain at least one Higgs singlet which can appear in odd powers.
The extended model with two doublets and one singlet is reviewed and
the general conditions which make the CP-violating $V_{min}$ the true
minimum are given. The generally called natural flavour-conserving models
are also discussed briefly.

 \endpage

\chapter{INTRODUCTION}

 The standard $SU(2)_L \times U(1)_Y$ electroweak theory
is a great success in describing the low-energy behaviors of the
electroweak interaction but fails to give successful
predictions on the fermion masses. It was first suggested
by $\rm Lee^1$ and then by many $\rm others^{2-6}$ that more Higgs
multiplets are needed to solve these problems. The most usual extension
of the standard model is to introduce another $SU(2)$ doublet.
It is possible to produce CP-violation spontaneously in such a model
without any additional symmetry. But we need some additional symmetry to
differentiate between these doublets.
In section 2.1, we show that if the doublets transform as $\Phi_1
\to \Phi_1$ and $\Phi_2 \to
e^{i\gamma}\Phi_2$, then no CP-violation can be produced
spontaneously.

  The two-Higgs doublet model is generally called the minimal extension
of the standard model. But it is not really the smallest one. The smallest
one should be that with one additional Higgs singlet only. It is also
the smallest model which allows the spontaneous CP-violation. Since
the Higgs singlet is allowed to appear in odd powers, not every phase in
the Higgs potential is to be annihilated by the symmetry introduced. We
may well say that it is the singlet $\chi$ which leads to the production of
spontaneous CP-violation (SCPV) in such models. But
it also leads to the baryon-number non-conservation if we require its Yukawa
couplings to the standard model fermions.

  We also investigate a larger $\rm extension^9$ with two Higgs doublets and
one singlet, in which a discrete $Z_3$ symmetry is used to distinguish the
multiplets. Such a model may spontaneously produce CP-violation in the
Higgs potential by suitably choosing the parameters. Calculations are made to
give the general conditions which make the CP-violating $V_{min}$ the true
minimum and some sets of suitably chosen parameters
are given in section 3.

  In addition to the discussions on
SCPV, we shall also discuss briefly
about the
traditional natural flavour-conserving (NFC) models and
mention a new natural model with an explicitly broken $S_3$ permutation
symmetry

 \chapter{THE CP-VIOLATION IN THE SMALLEST TWO EXTENSIONS}

\section{THE TWO-HIGGS DOUBLET MODEL}

  The standard electroweak model has only one Higgs doublet whose phase can
be rotated away by a gauge transformation so that no CP-violation would
appear spontaneously in the Higgs sector. It was widely $\rm suggested^{1-8}$
that one
needs at least two Higgs doublets to spontaneously produce CP-violation in the
Higgs sector. The two-Higgs doublet models generally have the FCNC problem
which arises from the non-simultaneous diagonalization of the two parts of the
fermion mass matrices corresponding to different doublets.
There are two traditional types of NFC models which
avoid the FCNC problem. One is to let the doublets couple to different types of
fermions seperately. The other is to let only one Higgs doublet couples to
all fermions. We also point out a type III NFC model with an
explicitly broken $S_3$ symmetry in Ref.[19]. In those three types of NFC
models mentioned above, there are no FCNC problem at tree level and it
remains true no matter what the Higgs masses are. It is interesting to know
whether SCPV exists in the Higgs potential in such models. We shall demonstrate
that any symmetry, under which the doublets transform as $\Phi_1 \to \Phi_1$
and $\Phi_2 \to e^{i\gamma}\Phi_2$,
would remove all SCPV terms from the potential. We note that the above
mentioned three NFC models all require a similar transformation on the scalar
fields .

  Without any additional symmety, the most general $SU(2)_L \times U(1)_Y$
gauge invariant Higgs potential in the two-Higgs doublet model
can be written as
$$\eqalign{ V= & \,\mu_1^2 \Phi_1^{\dagger} \Phi_1 +\mu_2^2 \Phi_2^{\dagger}
\Phi_2 +[\mu {\Phi_1^{\dagger} \Phi_2} + H.C.]
   +\lambda_1(\Phi_1^{\dagger} \Phi_1)^2 +\lambda_2(\Phi_2^{\dagger} \Phi_2)^2
\cr & \,+\lambda_3 (\Phi_1^{\dagger} \Phi_1)(\Phi_2^{\dagger} \Phi_2)
   +\lambda_4 (\Phi_1^{\dagger} \Phi_2)(\Phi_2^{\dagger} \Phi_1)
   +[{\lambda_5\over{2}} (\Phi_1^{\dagger} \Phi_2)^2 +H.C.] \cr & \,
   +[\lambda_6(\Phi_1^{\dagger} \Phi_1)(\Phi_1^{\dagger} \Phi_2)
+\lambda_7(\Phi_2^{\dagger} \Phi_2)(\Phi_1^{\dagger} \Phi_2) +H.C.] \cr
   = & \, V_0+V_{SCPV}}\eqno(2.1)$$
where $V_0$ is the CP-conserving part of the potential V, which contains
terms with $\Phi_i^{\dagger}$ and $\Phi_i$ appearing in pairs so as to
annihilate each other's phase. $V_{SCPV}$ contains terms
those may possess residual phases after spontaneous symmetry breaking (SSB).
We assume that all coefficients are real here.

 In order to break the $SU(2)_L \times U(1)_Y$ gauge symmetry down to
$U(1)_{EM}$ spontaneously, we have to choose the vacuum expectation
values of the Higgs fields as
$$<\Phi_1>={1\over{\sqrt{2}}} \pmatrix{0 \cr v_1},
   ~~~<\Phi_2>={1\over{\sqrt{2}}} \pmatrix{0 \cr v_2 e^{i\theta}} \eqno(2.2)$$
where we have used the gauge freedom to rotate away the phase of $<\Phi_1>$
and kept only the phase difference $\theta$ in $<\Phi_2>$.
Those terms may possess SCPV in the Higgs potential after SSB are picked
out as

 $$\eqalign{ V_{SCPV}= & \, \mu \Phi_1^{\dagger} \Phi_2
   +{\lambda_5\over{2}} (\Phi_1^{\dagger} \Phi_2)^2
   +\lambda_6(\Phi_1^{\dagger} \Phi_1)(\Phi_1^{\dagger} \Phi_2)
      +\lambda_7(\Phi_2^{\dagger} \Phi_2)(\Phi_1^{\dagger} \Phi_2)
     +H.C.} \eqno(2.3)$$

 Without any additional symmetry, we have seven unknowns ($\mu$,
$\lambda_5$, $\lambda_6$, $\lambda_7$, $v_1$, $v_2$ and $\theta$)
and only three equations from the minimization conditions,
${\partial{V_{SCPV}} \over {\partial x}} = 0$,
where $x=~v_1,~v_2~and ~\theta$.
It is impossible to determine the unknowns uniquely and
there are infinite sets of solutions.
We have the freedom to choose the parameters suitably to
satisfy the conditions and produce the SCPV in such a model.

  Consider certain symmetry under which the Higgs doublets transform as
$$ \Phi_1 \to \Phi_1,~~~~~~\Phi_2 \to e^{i\gamma} \Phi_2 \eqno(2.4)$$
where $\gamma$ is the angle depending on the chosen symmetry.
There are many discrete ($S_n$, $Z_n$, ...,etc) and continuous (PQ U(1) for
example) symmetries can give transformations like (2.4).
If we require the Lagrangian density be invariant under such symmetry,
then there are three ways to satisfy the requirement. The first way is to let
$\gamma=2n\pi$, n= integers. In this case, the doublets can be combined to
redefine a single Higgs doublet without losing
generality and it goes back to the standard model after all.
The second way is to let all the coefficients $\mu,~\lambda_5,~ \lambda_6$
and $\lambda_7$ vanish and the whole $V_{SCPV}$
will be gone. Both ways offer no SCPV. The third way is to
set $\gamma~=~\pi$ and make all $\mu,~
\lambda_6,~\lambda_7$ vanish, only the $\lambda_5$ term remains.

  Only the third way survives in the discussion above. But it will also
be proved to be CP-conserving. The proof is given in what follows.

  After spontaneous symmetry breaking, the Higgs potential would be written as
$$V~=~V_0~+~V_{SCPV}~=~V_0~+~{{v_1^2 v_2^2}\over 2}(\lambda_5 e^{2i\theta}
   ~+~\lambda_5^* e^{-2i\theta}) \eqno(2.5)$$
where we allow $\lambda_5$ to be complex for the moment.

  With the minimization condition on $\theta$, we find that
$${{\partial V} \over {\partial \theta}}~=~i v_1^2 v_2^2(\lambda_5
         e^{2i\theta}~-~ \lambda_5^* e^{-2i\theta})~=~0 \eqno(2.6)$$

If we define $\lambda_5~=~|\lambda_5|~e^{-i\alpha}$, then
${\lambda_5 / \lambda_5^*}~=~e^{-4i\theta}~=~e^{-2i\alpha}$.

 In case $\lambda_5$ is real, $\theta$ can be 0 or $\pi \over 2$.
For $\theta = 0$,
it goes back to the case discussed in Ref. [5]. For $\theta={\pi \over 2}$,
the vaccum expectation value of $\Phi_2$ would become
$<\Phi_2>={1\over{\sqrt{2}}} \pmatrix{0 \cr i v_2}$, which coincide with the
result obtained in our previous $\rm investegations^{7,8}$ considering an
$S_3$ discrete symmetry. Both cases produce no SCPV
in the Higgs potential. If $\lambda_5$ is complex, then $\alpha=2\theta$ and
all phases produced by the Higgs sector in $V_{SCPV}$ will be canceled out
by the phase of $\lambda_5$ so that no CP-violating phase would survive.

 We also consider the classification of those two doublets
$\Phi_1$ and$\Phi_2$ to a two-dimensional irreducible
representation
of symmetry like $S_3$. In this case, they transform as a doublet in such
symmetry and this doublet can be written as
$\Phi~=~\left(\matrix{\Phi_1 \cr \Phi_2}
\right)$ without losing generality.
The most general gauge invariant, renormalizable Higgs potential would be
given as
$$\eqalign{V= & \,a(\Phi_1^{\dagger},\Phi_2^{\dagger}) \left(\matrix{\Phi_1 \cr
\Phi_2}\right)
         +A[(\Phi_1^{\dagger},\Phi_2^{\dagger}) \left(\matrix{\Phi_1 \cr
\Phi_2}\right)]^2 \cr
            = & \, a(\Phi_1^{\dagger} \Phi_1 +\Phi_2^{\dagger} \Phi_2) +
A[(\Phi_1^{\dagger} \Phi_1)^2
         +(\Phi_2^{\dagger} \Phi_2)^2 +2(\Phi_1^{\dagger}
\Phi_1)(\Phi_2^{\dagger} \Phi_2)]}\eqno(2.7)$$
 In this potential, $\Phi_i^{\dagger}$ and $\Phi_i$ always appear in pairs
and no CP-violating phase would survive.

  We have considered several ways to construct the
Higgs potential in this
subsection and arrived at a conclusion on the production of the SCPV in
the two Higgs-doublet
model: If there is any symmetry under which the Higgs fields transform as in
(2.4), then no CP-violation could be produced spontaneously. The classification
of the doublets to a two-dimensional irreducible representation of some
symmetry also produces no SCPV.

\section{THE MINIMAL EXTENSION OF THE STANDARD MODEL}

 The truely minimal extension of the standard model is to add one singlet
$\chi$ rather than a doublet. We find that this singlet
dominates the production of the SCPV in such theories.

  Without any additional symmetry to constrain the Higgs fields,
the most general Higgs potential is

$$\eqalign{V= & \,a(\Phi^{\dagger} \Phi) + b(\chi^{\dagger} \chi) +
c(\Phi^{\dagger} \Phi)^2 + d(\chi^{\dagger} \chi)^2
   + e(\Phi^{\dagger} \Phi)(\chi^{\dagger} \chi) + V_{SCPV} \cr
   = & \, V_0 + V_{SCPV} \cr} \eqno(2.8)$$
where
$$\eqalign{V_{SCPV}= & \, A\chi^{\dagger} \chi^3 + B(\Phi^{\dagger}
\Phi)\chi^2
+ C \chi^4 + D\chi^3 + E\chi^{\dagger} \chi^2 + F(\Phi^{\dagger} \Phi)\chi +
G\chi^2 +h.c.}\eqno(2.9)$$
are those terms which may possess SCPV after SSB.
There are too many unknowns and too few equations
from the minimization conditions to give a unique set of
solutions. The general way to avoid this problem is to introduce some
symmetry to exclude some of the terms in $V_{SCPV}$ and reduce the number of
unknowns. If we introduce any symmetry under which the Higgs multiplets
transform as
$$\Phi \to \Phi~,~~~~~~~~\chi \to e^{i\gamma} \chi \eqno(2.10)$$
Then there are two ways to keep the potential invariant. One is to
exclude
those terms with coefficients E and F; the other one is to let $\gamma=2n\pi$,
n = integers, which is just an identity transformation. We still have
now too many unknowns and only three equations
to find a unique set of solutions. Just like what we have discussed in
the previous subsection, there are infinite sets of solutions
which may lead to the production of SCPV in the Higgs potential.

 If there is no other constraint on the potential
to make these parameters uniquely determined, then the one-doublet plus
one-singlet
$(\Phi~+~\chi)$ model is the smallest extension which
may produce CP-violation spontaneously in the Higgs potential.
But we meet the difficulty of baryon-number non-conservation in constructing a
gauge invariant Yukawa coupling. In the standard moodel,
since the physical left-handed fermions are classified to the $SU(2)$
doublet and the right-handed ones are to the $SU(2)$ singlet according to the
experimental data, we have to treat the Higgs field as
an $SU(2)$ doublet so as to have the baryon number conserved.
If we introduce an $SU(2)$ Higgs singlet into the theory and still want
it to couple with the standard model fermions, then we have to accept the SU(2)
singlet left-handed fermions  or the SU(2) doublet right-handed fermions to
conserve baryon numbers. Such classifications contradict the experiments.
In another way, we may introduce anti-fermion fields into the theory to
construct a gauge invariant Yukawa coupling. But it still does not conserve
baryon numbers. Neither ways are satisfactory to us.

\chapter{A LARGER EXTENSION WITH TWO DOUBLETS AND ONE SINGLET}

  A larger extension with two Higgs doublets and one singlet is discussed
in Ref. [9]. With a discrete $Z_3$ symmetry, it shows
that this model can produce CP-violation spontaneously in the Higgs potential
and the flavour conservation of the neutral currents is natural. But there
are some inaccurencies
in their calculation, which lead to a contrast conclusion on the minimal
potential.
Since the inaccurencies in Ref. [9], the set of parameters
chosen there would not make the CP-violating $V_{min}$ the true minimum as
is claimed. We demonstrate our
calculation to give the correct result and
express some sets of suitably chosen parameters which could make
the CP-violating $V_{min}$ the true minimum.

 In Ref. [9], an $SU(2)\times U(1)\times Z_3$ model with two
Higgs doublets and one singlet is suggested. The Higgs and fermion fields in
that model transform as
$$\phi_1\rightarrow exp[i(2m\pi /N)]\phi_1,~~~~~~~~~~\phi_2\rightarrow\phi_2,
{}~~~~~~~~\chi\rightarrow exp[i(2n\pi /N)]\chi .\eqno(3.1)$$
$$(u_R,N_R)\rightarrow exp[i(2m\pi /N)](u_R,N_R) . \eqno(3.2)$$
under $Z_N$ symmetry. Where the integers $m$ and $n$ obey $N > m,n > 0$ and
$N_R$ is used for the weak $I_z={-1\over 2}$ fermions.
With N=3, the most general gauge invariant Higgs potential is given as
$$\eqalign{V= & \,m_1^2 \Phi_1^{\dagger} \phi_1 + m_2^2\Phi_2^{\dagger} \phi_2
+m_3^2 \chi^{\dagger} \chi +
a_{11} (\Phi_1^{\dagger} \phi_1 )^2 +a_{22}(\Phi_2^{\dagger} \phi_2 )^2
+a_{33}(\chi^{\dagger} \chi)^2 \cr + & \,
a_{12}(\Phi_1^{\dagger} \phi_1)(\Phi_2^{\dagger} \phi_2) +
a_{13}(\Phi_1^{\dagger} \phi_1)(\chi^{\dagger} \chi)
+a_{23}(\Phi_2^{\dagger} \phi_2)(\chi^{\dagger} \chi) + b_{12}(\Phi_1^{\dagger}
\phi_2)(\Phi_2^{\dagger} \phi_1) +V_{SCPV}\cr = &\
 V_0 +V_{SCPV}\cr} \eqno(3.3) $$
where $V_0$ and $V_{SCPV}$ are, respectively, the CP-conserving and
spontaneously CP-violating parts of the Higgs potential. The Yukawa
couplings are written ${\rm as^{10}}$
$$\eqalign{L_Y = & \,h^u {\bar q}_L {\tilde \phi}_1 u_R + h^d {\bar q}_L \phi_2
     +h^e {\bar l}_L \phi_2 e_R +h^{\nu} {\bar l}_L {\tilde \phi}_1 N_R
      +h.c.+L_{LNV} \cr \equiv & \,L_{Y_0} +L_{LNV}}.\eqno(3.4)$$

The details of $V_{SCPV}$ and $L_{LNV}$ depend on the chosen discrete symmetry.
The discrete symmetry considered here is
$Z_3$ symmetry. Under this symmetry, $V_{SCPV}$ and
$L_{LNV}$ take the following form
$$V_{SCPV}= c_1 (\Phi_1^{\dagger} \phi_2)\chi^{\dagger} \chi^{\dagger} + c_2
(\Phi_1^{\dagger} \phi_2)\chi + c_3 \chi^3 + h.c.,\eqno(3.5)$$
$$L_{LNV}=h^N~ N^T_R ~ N_R~ \chi +h.c.\eqno(3.6)$$

We note that for paticular sets of (m, n) = (1, 1) or (2, 2), those three terms
in (3.5) are $Z_3$ invariant.
The vacuum expectation values of the Higgs multiplets are chosen as
$$<\phi_1>={v_1 exp(i\theta)\over \sqrt2},~~~~~~~~<\phi_2>={v_2\over \sqrt2},
{}~~~~~~~~<\chi>={v_3  exp(i\alpha)\over\sqrt2} \eqno(3.7)$$
where the phase of $<\phi_2>$ has been rotated away and the phase differences
between these multiplets
are expressed as $\theta$ in $<\Phi_1>$ and $\alpha$ in $<\chi>$ as in Ref.
[9].
After SSB, the minimal potential V with CP conserved
is given by the condition
$$2\alpha + \theta = n \pi,~~~~\alpha - \theta = m\pi \eqno(3.8)$$
where n and m are integers. While that with CP-violated is given by the
conditions
$$cos(2\alpha +\theta)={{c_2 c_3}\over {c_1^2 v_1 v_2}}-{{c_3 v_3^2}\over {2c_2
    -{c_2 v_1 v_2\over 2c_3 v_3^2}$$
$$cos(\alpha -\theta)={\sqrt{2}c_1 c_3 v_3^3\over 4c_2^2 v_1 v_2}
 -{\sqrt{2}c_3 v_3\over 2c_1 v_1 v_2}-{{\sqrt{2}c_1 v_1 v_2}\over {4c_3
        v_3}}\eqno(3.9)$$

The minima of the potential for these two cases are given in Ref. [9] as

CP-conserving:
$$V_{min}^{CP}=V_0 +{1\over2} (c_1 + \sqrt{2} c_2)v_1 v_2 v_3^2 +{c_3 v_3^2
\over \sqrt{2}}\eqno(3.10)$$

and CP-violating:
$$V_{min}=V_0-{c_2 c_3 v_3^2\over4c_1} -{c_1 c_3 v_3^4\over4c_2}
 -{c_1 c_2 v_1 v_2\over4c_3}\eqno(3.11)$$

But we find that these two expressions (3.10) and (3.11) should be corrected
to those in (3.14) and (3.15) as you shall see.
It is claimed that $V_{min}$ will be the absolute minimum if one chooses
a set of parameters suitablly as
$$c_2 \sim c_3 \sim  -c_1 v_3 > 0,~~~~~~~~v_3^2 \sim v_1 v_2\eqno(3.12)$$
then one should arrives at
$ \alpha = \pi /4$ and $\theta =0$.
The relation between the CP-violating and the CP-conserving potentials is
shown as
$$V_{min}=V_0+{3\over 4} \vert {c_1} \vert v_1^2 v_2^2 < V_{min}^{CP}=V_0
 +{1\over 2}(2 \sqrt{2} -1) \vert {c_1} \vert v_1^2 v_2^2 \eqno(3.13)$$

  Since the equations (3.10) and (3.11) are incorrect,
this inequality should be corrected to (3.18). In our
calculation, we find that (3.10) and (3.11) should be corrected to:

CP-conserving (corrected):
$$V_{min}^{CP}=V_0+{c_1 v_1 v_2 v_3^2\over2}+{c_2 v_1 v_2 v_3 \over \sqrt2}
+{c_3 v_3^3\over \sqrt2}\eqno(3.14)$$
where the orders of $v_3$ in the third and fourth terms have been corrected,

and CP-violating (corrected):
$$V_{min}=V_0-{c_2 c_3 v_3^2\over 2c_1} -{c_1 c_3 v_3^4\over 4c_2} -{c_1 c_2
v_1^2 v_2^2\over 4c_3}.\eqno(3.15)$$
where the coefficient of the second term is 1/2 rather than 1/4 and
the orders of $v_1 v_2$ in the fourth term should be of order two rather than
one. If one follows the choice of parameters in (3.12), then one would arrive
at the corrected minima
$$V_{min}=V_0+ \vert {c_1} \vert v_1^2 v_2^2 \eqno(3.16)$$
and
$$V_{min}^{CP}=V_0+{1\over 2} (2 \sqrt{2} -1) \vert {c_1} \vert c_1^2 v_2^2
\eqno(3.17)$$
where the coefficient of the second term in $V_{min}$ should be 1 rather than
$3/4$. Therefore, we find that the relation between $V_{min}$ and
$V^{CP}_{min}$
is inverse to (3.13) and should be corrected to

$$ V_{min}=V_0+\vert {c_1} \vert v_1^2 v_2^2 > V_{min}^{CP} = V_0+
{(2\sqrt{2}-1)\over2}\vert {c_1} \vert v_1^2 v_2^2.\eqno(3.18)$$
The true absolute minimum under the choice of (3.12) is $V^{CP}_{min}$
rather than $V_{min}$.
But the choice of parameters is somewhat arbitrary, so we may choose other
sets of parameters suitably to make $V_{min}$ the true minimum.
For instance, one may choose a set of parameters as what follows and avoid the
hierachy problem in the Higgs potential.
$$c_1 \sim c_2 ,~~~~~~~~ v_3^2 \sim v_1 v_2 , ~~~~~~~~c_3~>~0 \eqno(3.19)$$
If we require $V^{CP}_{min}~>~V_{min}$, i.e, $V_{min}$ to be the true minimum,
then we must have $c_3~>~0$ since this requirement gives an inequality

$$ {1 \over{c_3}} [v_3 (c_1 +c_3)+ \sqrt{2} c_3]^2~>~0 \eqno(3.20)$$
Besides this set of choice we find that if we choose any two of $\rm c_1, c_2
{}~and~c_3$ to be equal and $v_3^2 \sim v_1 v_2$, then the other $c_i >0$ will
make $V_{min}$ the true minimum. So we have too much freedom to choose the
parameters uniquely.

\chapter{SUMMARY AND DISCUSSIONS}

  In this article, we have discussed the SCPV in three kinds of extension
of the standard model. We find
that if there is no additional symmetry introduced to distinguish the Higgs
multiplets, then every model possess SCPV. When symmetry is introduced, under
which the multiplets transform as (2.4), the two-Higgs doublet model will
lose its SCPV terms. While the smallest extension of the standard model with
one Higgs
singlet added only still posses the SCPV terms even when the symmetry
under which the Higgs multiplets transform as in (2.10) is introduced. But it
leads to baryon number non-conservation.

  In the larger extended model with two Higgs doublets and one singlet, in
which
a discrete $Z_3$ symmetry is imposed, there are too
many unknown parameters and too few equations to find a unique set of
solutions. But we can choose the parameters suitably to make the
CP-violating minimum $V_{min}$ the true minimum. Calculations are demonstrated
t
correct the previous mistakes in Ref. [9] and the general conditions for
the spontaneous CP violation are also given.

  We note that the Higgs singlet plays an important role in producing SCPV in
the Higgs potential. Since the Higgs singlet can appear in odd powers,
there must be terms whose phases are not annihilated after SSB
even when additional symmetry is introduced.
If one only cares about the SCPV, then the Higgs singlet alone is enough. But
if one wants to preserve the baryon numbers, then the Higgs singlet can not
have Yukawa couplings.

  The two-Higgs doublet models generally have the
FCNC problem which arises from the non-simultaneous diagonalization
of the two parts of the fermion mass matrices corresponding to
different Higgs doublets.
There are two traditional NFC
models to $avoid$ this problem. One is to let only one of the doublets
couples
to both types of fermions and the other to none. The other way is to let those
t
doublets couple
to different types of fermions seperately.
We have pointed out another
way (the type III NFC $\rm model^{19}$) to suppress
the FCNC's naturally with an explicitly broken $S_3$ symmetry.
In these three models, there are no FCNC problems and also have no SCPV terms
in the Higgs potential as proved in section 2.1.

\ref{T.D. Lee, Phys. Rev. D8, 1226(1973); Phys. Rep. 9, 143(1974).}
\ref{H.E. Haber, G.L. Kane, and T. Starling, Nucl.Phys. B161, 493(1979).}
\ref{K.S. Babu, and Ernest Ma, Phys. Rev. D31 2861(1985).}
\ref{S. Bertolini and A. Sirlin, Nucl. Phys. B248, 589(1984).}
\ref{S. Bertolini, Nucl. Phys. B272, 77(1986).}
\ref{G.C. Branco, A.J. Buras and J.M. Gerard, Nucl. Phys. B259, 306(1985).}
\ref{Chien-er Lee, Chilong Lin, and Yeou-Wei Yang, Chin. J. Phys. 26,
180(1988).}
\ref{Chien-er Lee, Chilong Lin and Yeou-Wei Yang Phys. Rev. D42, 2355(1990).}
\ref{C.Q. Geng and J.N. Ng Phys. Lett. B211, 111(1988).}
\ref{N.G. Deshpande, M. Gupta,and P.B. Pal, Phys. Rev. D45. 953(1992).}
\ref{E. Ma, Phys. Rev. D43. R2761(1991).}
\ref{E. Ma, Phys. Rev. D44. R587(1991).}
\ref{D. Wyler, Phys. Rev. D19. 3369(1979).}
\ref{E. Derman, and H.S. Tsao, Phys. Rev. D20. 1207(1979).}
\ref{H. Ozaki, Phys. Rev. D40. 2425(1989).}
\ref{H.Fritzsch, and J. Plankl, Phys. Rev. D43. 3026(1991).}
\ref{Chilong Lin, Chien-er Lee, and Yeou-Wei Yang, NCKU-TH-HEP, July, (1991).}
\ref{Chien-er Lee, Yeou-Wei Yang, Y-L. Chang, and S-N. Lai, Chin. J. Phys. 24,
2
\ref{Chilong Lin, Chien-Er Lee, and Yeou-Wei Yang, to be
              published in Physical Review D}
\refout
\end